\documentclass[sigconf]{acmart}
\usepackage{subcaption}
\usepackage[most]{tcolorbox}
\usepackage{soul}
\usepackage{color, xcolor}

\renewcommand\footnotetextcopyrightpermission[1]{} 
\acmConference[Workshop '25]{xxxxxxx}{xxxxxx, 2025}{xxx, xx}
\acmPrice{15.00}
\acmISBN{978-1-4503-XXXX-X/18/06}

\begin{document}

\title{On The Reproducibility Limitations of RAG Systems}

\author{Baiqiang Wang}
\affiliation{%
  \institution{University of Washington}
  \city{Seattle}
  \state{WA}
  \country{USA}
}
\email{wbq@uw.edu}

\author{Dongfang Zhao}
\affiliation{%
  \institution{University of Washington}
  \city{Seattle}
  \state{WA}
  \country{USA}
}
\email{dzhao@uw.edu}

\author{Nathan R Tallent}
\affiliation{%
  \institution{Pacific Northwest National Laboratory}
  \city{Richland}
  \state{WA}
  \country{USA}
}
\email{nathan.tallent@pnnl.gov}

\author{Luanzheng Guo}
\affiliation{%
  \institution{Pacific Northwest National Laboratory}
  \city{Richland}
  \state{WA}
  \country{USA}
}
\email{lenny.guo@pnnl.gov}
\begin{abstract}
Retrieval-Augmented Generation (RAG) is increasingly employed in generative AI-driven scientific workflows to integrate rapidly evolving scientific knowledge bases, yet its reliability is frequently compromised by non-determinism in their retrieval components. This paper introduces \textbf{ReproRAG}, a comprehensive benchmarking framework designed to systematically measure and quantify the reproducibility of vector-based retrieval systems. ReproRAG investigates sources of uncertainty across the entire pipeline, including different embedding models, precision, retrieval algorithms , hardware configurations, and distributed execution environments. Utilizing a suite of metrics, such as Exact Match Rate, Jaccard Similarity, and Kendall’s Tau, the proposed framework effectively characterizes the trade-offs between reproducibility and performance. Our large-scale empirical study reveals critical insights; for instance, we observe that different embedding models have remarkable impact on RAG reproducibility. The open-sourced ReproRAG framework provides researchers and engineers productive tools to validate deployments, benchmark reproducibility, and make informed design decisions, thereby fostering more trustworthy AI for science.
\end{abstract}




\maketitle

\section{Introduction}

Retrieval-Augmented Generation (RAG)~\cite{lewis2020rag} has become a cornerstone architecture for building knowledge-intensive AI systems, particularly in scientific research~\cite{llm_for_science_survey_2024}. By grounding LLMs with evidence retrieved from vast corpora, RAG systems mitigate factual hallucinations~\cite{ji2023survey_hallucination} and provide access to domain-specific, timely information. However, as these systems move from prototypes to critical components in scientific workflows, their reliability and consistency become paramount, a central concern in the broader pursuit of Trustworthy AI~\cite{sun2024trustllm}.

A fundamental challenge that threatens this reliability is non-determinism, a well-documented issue in parallel and deep learning systems~\cite{nagarajan2018irreproducibility}. A RAG system, even with identical inputs, can produce different results across multiple runs. This inconsistency often originates in the vector-based retrieval stage. Despite its importance, there is a lack of standardized tools and methodologies to quantify this specific form of non-determinism. While performance benchmarking for vector search has been studied~\cite{ann_benchmarks}, it does not address reproducibility, forcing researchers to rely on anecdotal evidence or ad-hoc tests. Without rigorous measurement, it is impossible to understand the trade-offs between performance and consistency, validate production deployments, or confidently build upon prior computational results.

To fill this measurement gap, we introduce \textbf{ReproRAG}, a comprehensive testing framework for analyzing and measuring the reproducibility of RAG retrieval systems. Our contributions are:
\begin{itemize}
    \item We design and implement \textbf{ReproRAG}, an open-source framework for systematically benchmarking the reproducibility of RAG retrieval pipelines, defining a suite of metrics to quantify multiple dimensions of uncertainty.
    \item We present a large-scale empirical study that establishes a \textbf{hierarchy of uncertainty}, demonstrating that the choice of embedding model and dynamic data insertion are the most significant sources of result variation.
    \item We provide a detailed empirical analysis that challenges common assumptions, demonstrating that core ANN retrieval algorithms and distributed protocols can achieve perfect run-to-run reproducibility, and that common software-level determinism flags may have no observable effect on modern embedding models.
    \item We offer \textbf{actionable insights for practitioners}, including a quantification of precision-induced "embedding drift" and evidence that performance-oriented settings can provide a speedup without harming reproducibility for certain classes of models.
\end{itemize}

\section{Background and Related Work}
\label{sec:background}

\subsection{Vector Search and Non-Determinism}
Modern RAG systems rely on Approximate Nearest Neighbor (ANN) search to efficiently find relevant documents in high-dimensional vector spaces. Libraries like FAISS (Facebook AI Similarity Search)~\cite{johnson2021faiss} implement various ANN algorithms. These methods often trade perfect accuracy for significant speed improvements, but this trade-off can introduce sources of non-determinism.

For instance, \textbf{Hierarchical Navigable Small World (HNSW)}~\cite{malkov2018hnsw} builds a probabilistic graph for searching, where choices made during graph construction can be a source of randomness. Similarly, \textbf{Inverted File (IVF)} indexes first partition the vector space using k-means clustering, whose random initialization can lead to different index structures~\cite{jegou2011product}. Another common technique, \textbf{Locality-Sensitive Hashing (LSH)}~\cite{gionis1999similarity}, uses random projections to group similar items, where the choice of projections is inherently stochastic. Furthermore, beyond algorithmic randomness, parallel execution on multi-core CPUs or GPUs can introduce non-determinism through the reordering of floating-point operations.

\subsection{Related Work}
Our work is situated at the intersection of three key areas: the application of RAG in science, the broader challenge of computational reproducibility, and the practice of benchmarking AI systems.

\paragraph{RAG in Scientific Workflows}
The application of Large Language Models to accelerate scientific discovery is a rapidly growing field~\cite{llm_for_science_survey_2024}. RAG frameworks, in particular, are critical for grounding these models in specialized, up-to-date scientific literature, making them powerful tools for tasks like hypothesis generation and literature review. While frameworks like SciTrust~\cite{herron2024scitrust} evaluate the trustworthiness of the LLM component itself, they assume a stable, reproducible input. Our work is complementary, focusing on ensuring the reliability of the retrieval stage that produces this input.

\paragraph{The Reproducibility Crisis in Computational Science}
The challenge of reproducibility is a well-documented issue across computational science~\cite{donoho2015}. In response, the high-performance computing community has championed efforts to improve the repeatability of research, for instance, through the introduction of Artifact Description and Evaluation tracks at major conferences~\cite{collberg2014}. Our framework applies these established principles of rigorous measurement and validation to the specific and novel challenges posed by large-scale vector search systems.

\paragraph{Benchmarking AI and Vector Search Systems}
Standardized benchmarking is essential for the progress of AI systems. Efforts like MLPerf~\cite{MLPerf} have become industry standards for measuring ML training and inference performance. In the domain of vector search, frameworks like \texttt{ann-benchmarks}~\cite{ann_benchmarks} provide comprehensive performance comparisons of different ANN algorithms. ReproRAG extends this tradition of rigorous benchmarking to a new and critical dimension: reproducibility. While existing benchmarks focus on speed and accuracy, our work provides the first comprehensive framework for quantifying the consistency and deterministic behavior of these systems.

\section{The Test Framework}
\label{sec:framework}

ReproRAG is architected to be a flexible and scalable testing harness. It systematically creates experimental configurations, executes repeated retrieval runs, and analyzes the results to generate detailed reproducibility reports.

\subsection{Sources of Non-Determinism Investigated}
The framework is designed to isolate and measure uncertainty from four primary sources:
\begin{enumerate}
    \item \textbf{Embedding Uncertainty:} Variations arising from the choice of embedding models and their numerical precision (e.g., \texttt{FP32} vs. \texttt{FP16}).
    \item \textbf{Retrieval Uncertainty:} Inconsistencies stemming from the choice of indexing techniques (e.g., Flat, IVF, HNSW) and the selection of parameters.
    \item \textbf{Hardware Variations:} The impact of the underlying hardware, including differences between CPU and GPU execution and parallel execution effects.
    \item \textbf{Distributed Computing:} Consistency challenges in multi-node environments, including data sharding strategies and inter-node communication.
\end{enumerate}

\subsection{Architecture and Methodology}
Motivated by these challenges, ReproRAG employs a systematic methodology integrating multi-level testing and controls.

\paragraph{System Architecture} The framework is built with a modular architecture: an \textbf{ExperimentConfig} module for defining test parameters; a \textbf{FaissRetrieval} core engine; an MPI-based \textbf{DistributedFaissRetrieval} module; and a \textbf{ReproducibilityMetrics} library for analysis.

\paragraph{Determinism Control} To isolate sources of randomness, the framework implements comprehensive controls, including standardization of random seeds, enforcement of deterministic CUDA operations, and fixed thread counts.

\paragraph{Multi-Precision Analysis} To evaluate embedding uncertainty, the framework tests multiple precision configurations, including full-precision \texttt{FP32}, and half-precision \texttt{FP16} for memory efficiency. It also supports specialized formats like \texttt{TF32} and \texttt{BF16} on compatible hardware.

\paragraph{Hardware and Environment Profiling} The framework automatically detects GPU architecture, profiles memory utilization, and logs driver and library versions to ensure environmental consistency is tracked throughout all experiments.

\subsection{Core Reproducibility Metrics}
ReproRAG utilizes two levels of metrics.

\paragraph{Embedding Stability Metrics}
To measure the consistency of the generated vectors directly, before they are used in a search index, we compute:
\begin{itemize}
    \item \textbf{L2 Distance}~\cite{leskovec2014mmds}: Also known as Euclidean distance, this measures the absolute, straight-line distance between two embedding vectors in the vector space. A score of 0.0 indicates that the vectors are numerically identical.
    \item \textbf{Cosine Similarity~\cite{manning2008ir}}: This measures the cosine of the angle between two vectors, evaluating their semantic orientation independent of magnitude. A score of 1.0 indicates the vectors point in the exact same direction.
\end{itemize}

\paragraph{Retrieval Metrics} To measure the final outcome of the retrieval task, we use:
\begin{itemize}
    \item \textbf{Jaccard Similarity~\cite{leskovec2014mmds}}: This metric evaluates the overlap between two sets of retrieved documents (A and B), ignoring their rank. It is calculated as the size of the intersection divided by the size of the union ($|A \cap B| / |A \cup B|$). A score of 1.0 indicates the two runs returned the exact same set of documents. 
    \item \textbf{Kendall's Tau~\cite{kendall1970rank}:} This measures the rank correlation between two lists of retrieved results. It evaluates the similarity of the ordering of documents by counting the number of concordant and discordant pairs. A score of 1.0 indicates that the documents common to both lists were ranked in the exact same order. 
    \item \textbf{Score Stability:} This is the standard deviation of the similarity scores (e.g., L2 distances) of the top-k retrieved documents across multiple runs. A lower value indicates higher numerical consistency in the ranking function itself.
\end{itemize}

\section{Experimental Evaluation}
\label{sec:evaluation}

We now present our empirical evaluation using the ReproRAG framework to characterize reproducibility across several key dimensions discussed in Section~\ref{sec:framework}.

\subsection{Experimental Setup}
Unless otherwise specified, all experiments share the following setup:
\begin{itemize}
    \item \textbf{Platform:} The TAMU ACES HPC platform~\cite{aces_platform}, utilizing both CPU-only and GPU-accelerated nodes.
    \item \textbf{Dataset:} We use a Medium Scale configuration (100,000 documents; 1,000 queries) from MSMARCO dataset~\cite{msmarco}.
    \item \textbf{Embedding Models:} To ensure our findings are generalizable, we evaluated three diverse, high-performing, and widely-used transformer-based models:
    \begin{itemize}
        \item \textbf{BGE (BAAI General Embedding)}~\cite{bge_model}: We used the \texttt{bge-base-en-v1.5} model, a powerful sentence embedding model known for its strong performance on retrieval benchmarks.
        \item \textbf{E5 (Embedding from ELI5)}~\cite{wang2022e5}: We used the \texttt{intfloat/e5\\-base-v2} model, which is notable for its effective training on a massive corpus of text pairs using contrastive pre-training.
        \item \textbf{Qwen}~\cite{qwen3technicalreport}: We used the \texttt{Qwen3-0.6B-Base} model, a recent and powerful open-source text embedding model from a new generation of large language models.
    \end{itemize}
Our study characterizes the reproducibility of these specific models, which represent different training methodologies and architectural families.
    \item \textbf{Measure:} For each configuration, every query is executed over five runs. We report the average measurement.
\end{itemize}

\subsection{Scenario 1: Temporal Reproducibility Under Data Insertion}
\subsubsection{Goal and Methodology}
Our evaluation begins by addressing a practical challenge that extends beyond static, run-to-run consistency: \textbf{temporal reproducibility}. We aim to answer the question: how consistently can an index reproduce its initial search results after new data has been incrementally added to it? This scenario evaluates the stability of search outcomes over time in a common real-world situation where rebuilding an index from scratch after every update is computationally prohibitive.

\paragraph{Experimental Protocol}
The methodology is designed to quantify the reproducibility of search results between a ``before'' (\texttt{V1}) and ``after'' (\texttt{V2}) state of the same index instance.
\begin{enumerate}
    \item \textbf{Initial State (V1):} An index (e.g., HNSW, IVF) is created and trained (if applicable) using an initial dataset of \textasciitilde7,920 documents. This establishes a fixed internal structure. These initial documents are then added to the index.
    \item \textbf{Baseline Query:} A standard set of 100 queries is executed against this V1 index. The top-50 retrieved document IDs for each query are recorded as \texttt{Results\_V1}, representing the original, reproducible state.
    \item \textbf{Incremental Update:} A new set of \textasciitilde1,980 documents is added directly to the same index instance using the \texttt{index.add()} function. No re-training or re-structuring of the index occurs.
    \item \textbf{Follow-up Query:} The exact same 100 queries are executed again against the updated index. The top-50 results are recorded as \texttt{Results\_V2}.
\end{enumerate}

\paragraph{Stability Metrics}
The degree to which \texttt{Results\_V1} are reproduced in \texttt{Results\_V2} is quantified using metrics specifically suited for comparing ranked lists:
\begin{itemize}
    \item \textbf{Overlap Coefficient:} The percentage of documents from the V1 result list that are also present in the V2 list.
    \item \textbf{Kendall's Tau:} Measures the rank-order correlation for the documents that appear in both result lists.
    \item \textbf{Rank-Biased Overlap (RBO)\cite{webber2010similarity}:} A comprehensive, top-weighted score that evaluates the similarity of the two lists, penalizing disagreements at higher ranks more severely. An RBO of 1 indicates perfect temporal reproducibility.
\end{itemize}

\subsubsection{Results and Discussion}
Our experiment reveals a nuanced and consistent picture of how different indexing methods handle incremental updates. The primary finding is that while data insertion inevitably causes some ``churn'' in the retrieved results, the internal ranking of the original, persistent results remains perfectly stable across all tested approximate indexing methods.

\paragraph{Quantifying Result Churn}
Table~\ref{tab:insertion_stability} summarizes the stability metrics, averaged across 100 queries. All three methods—HNSW, IVF, and LSH—exhibited very similar behavior. The mean Overlap Coefficient was approximately 0.80 for all methods, indicating that on average, 80\% of the original top-50 documents were still present in the results after the data insertion. The Rank-Biased Overlap (RBO) score, which gives more weight to top-ranked items, was also remarkably consistent across the methods, averaging around 0.73.

\begin{figure}[ht!]
    \centering
    \begin{subfigure}[b]{0.48\columnwidth}
        \centering
        \includegraphics[width=\textwidth]{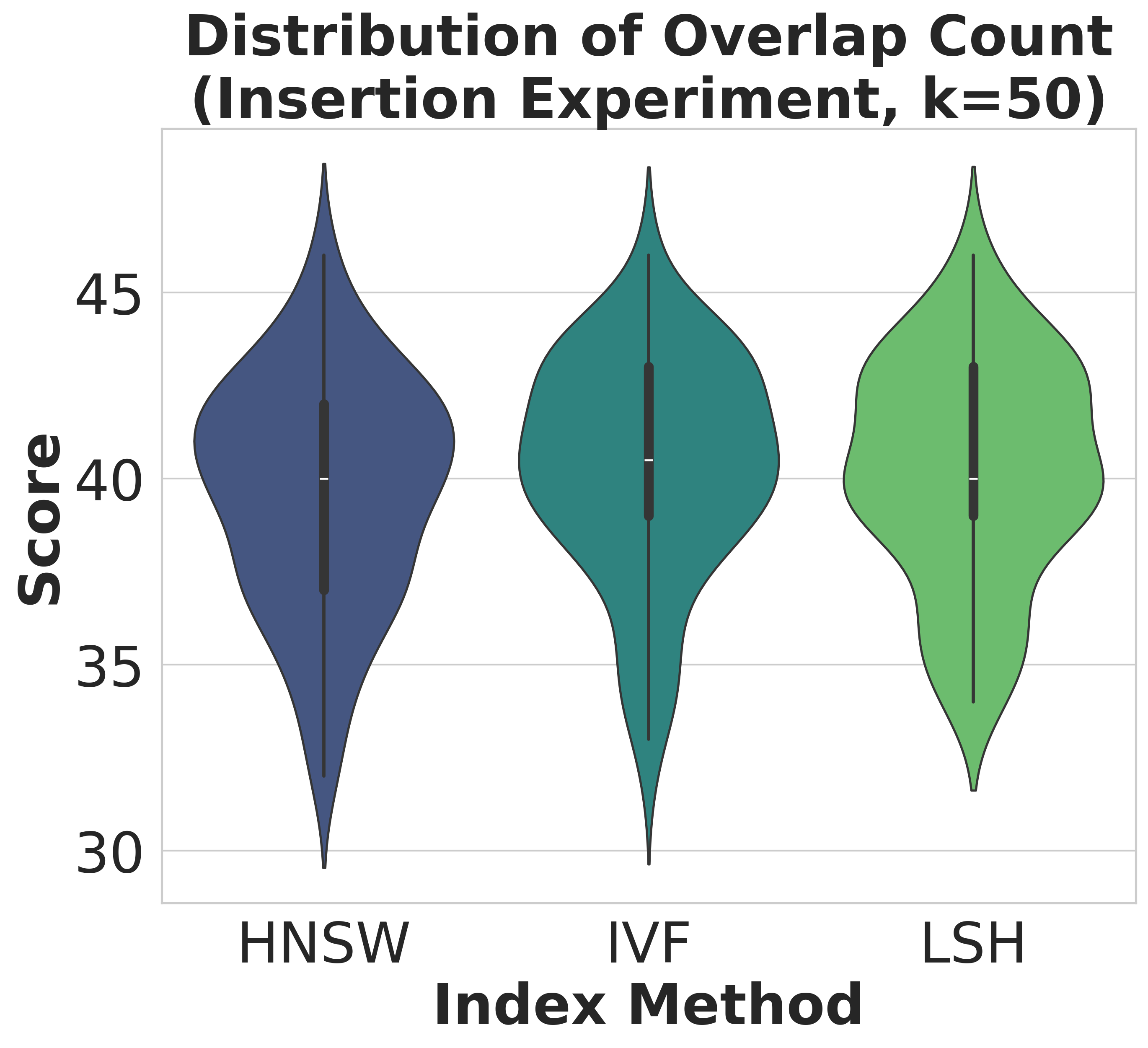}
        \caption{Overlap Count}
        \label{fig:overlap_count_dist}
    \end{subfigure}
    \hfill
    \begin{subfigure}[b]{0.48\columnwidth}
        \centering
        \includegraphics[width=\textwidth]{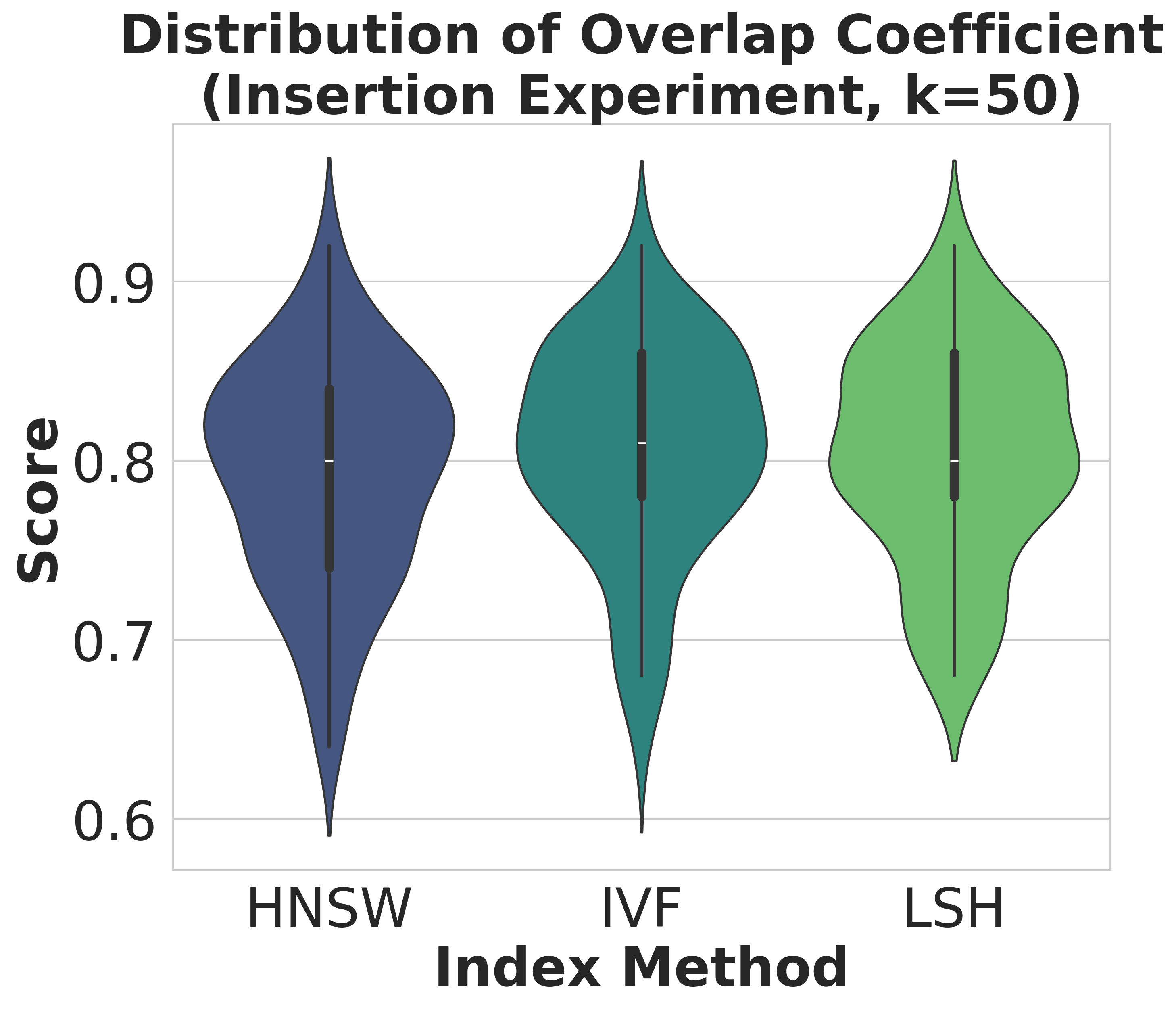}
        \caption{Overlap Coefficient}
        \label{fig:overlap_coeff_dist}
    \end{subfigure}
    
    \vspace{1em} 
    
    \begin{subfigure}[b]{0.48\columnwidth}
        \centering
        \includegraphics[width=\textwidth]{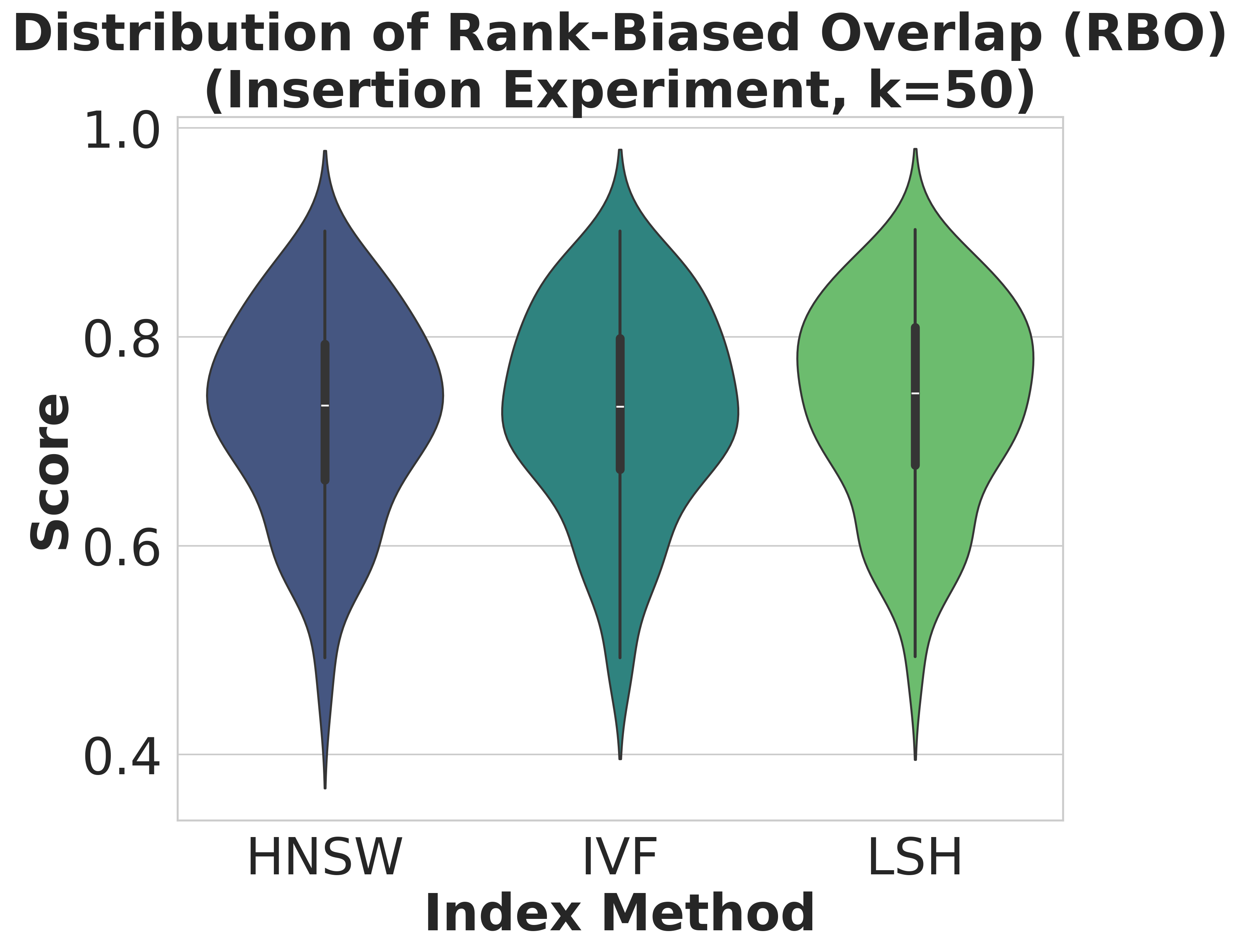}
        \caption{Rank-Biased Overlap (RBO)}
        \label{fig:rbo_dist}
    \end{subfigure}
    \hfill
    \begin{subfigure}[b]{0.48\columnwidth}
        \centering
        \includegraphics[width=\textwidth]{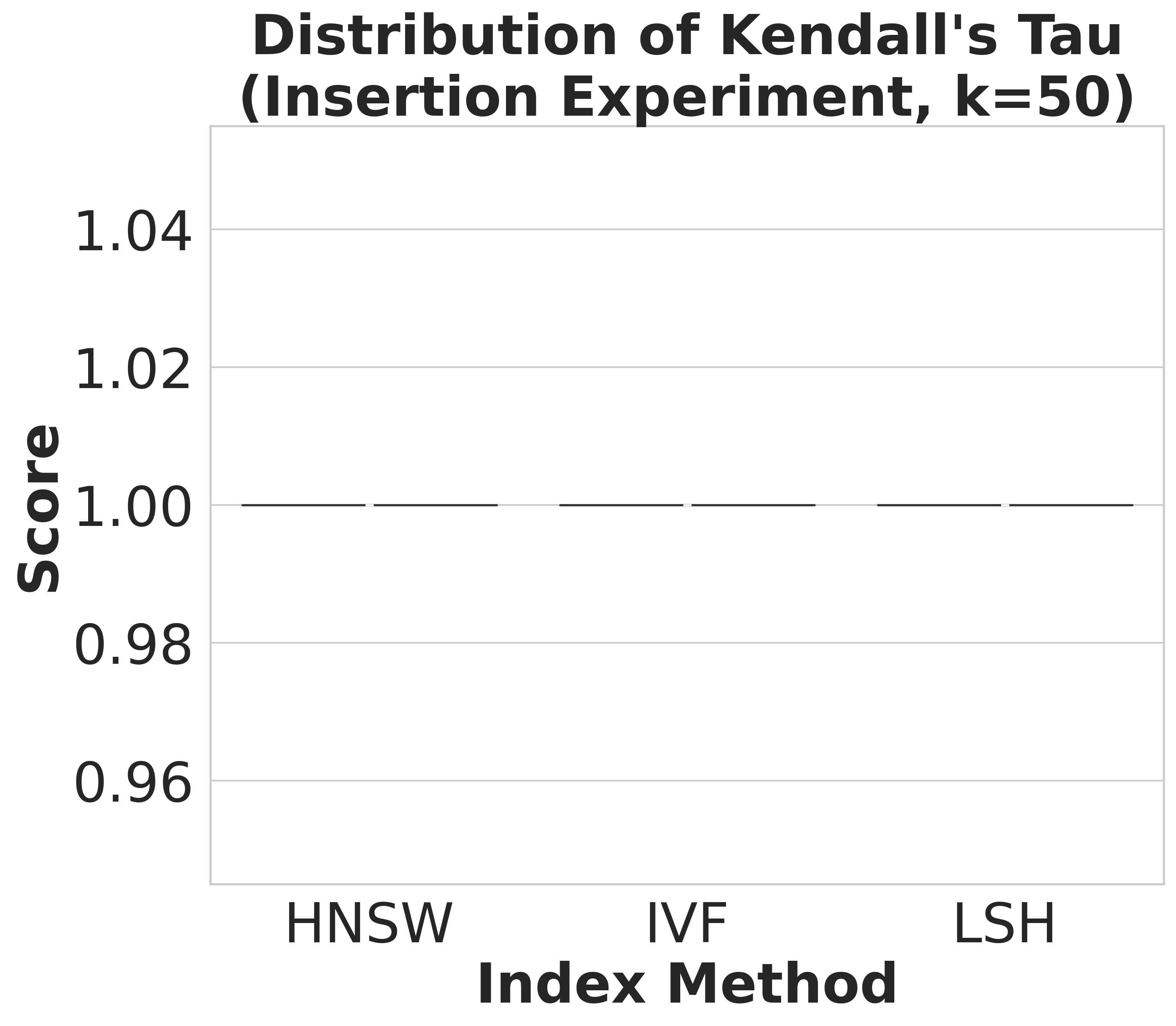}
        \caption{Kendall's Tau}
        \label{fig:kendall_dist}
    \end{subfigure}
    \caption{Distributions of temporal reproducibility metrics for the HNSW index across 100 queries after data insertion. The results show variance in overlap (a, b) and RBO (c), but perfect stability in ranking (d).}
    \label{fig:insertion_distributions}
\end{figure}

Figure~\ref{fig:insertion_distributions} visualizes the distribution of these stability metrics across the 100 test queries for the HNSW index. While the mean scores in Table~\ref{tab:insertion_stability} provide a good summary, the distributions show there is notable variance depending on the specific query. For instance, the Overlap Coefficient (Figure~\ref{fig:overlap_coeff_dist}) has a median of 0.80 but ranges from 0.64 to 0.92, indicating that some queries are much more sensitive to data insertion than others. The RBO distribution (Figure~\ref{fig:rbo_dist}) shows a similar spread.

\begin{table}[h]
  \caption{Retrieval Stability After Incremental Data Insertion. Metrics are the mean values across 100 queries.}
  \label{tab:insertion_stability}
  \begin{tabular}{lccc}
    \toprule
    Index Type & Overlap Coeff. & RBO Score & Kendall's Tau \\
    \midrule
    \texttt{HNSW} & 0.793 & 0.725 & 1.000 \\
    \texttt{IVF} & 0.806 & 0.730 & 1.000 \\
    \texttt{LSH} & 0.804 & 0.735 & 1.000 \\
    \bottomrule
  \end{tabular}
\end{table}

\paragraph{Perfect Ranking Stability of Persistent Results}
The most striking result, visualized in Figure~\ref{fig:kendall_dist}, is the perfect Kendall's Tau score of 1.000 (with zero standard deviation) for all three index types. This reveals a critical insight into the nature of the result churn. It means that for the subset of documents that appeared in both the ``before'' and ``after'' result lists, their relative ranking did not change at all. The instability we measured is not due to a re-shuffling of the original results, but rather a process of **displacement**, where new or newly-promoted documents push some of the original results out of the top-50 list.

\paragraph{Discussion}
\textcolor{blue}{The key takeaway from this scenario is that for these common indexing methods, the primary effect of ``lazy'' data insertion is displacement, not re-ranking.} This makes their behavior more predictable than might be assumed. The perfect Kendall's Tau suggests that the underlying similarity landscape for the original documents is not fundamentally distorted by the addition of new data points. This is a crucial piece of information for practitioners building systems for dynamic environments. It implies that users are likely to see many of their original results preserved, with new, potentially more relevant, items added to the list, rather than seeing a chaotic re-ordering of familiar results.

\subsection{Scenario 2: Cross-Embedding Model Retrieval Consistency}
\subsubsection{Goal and Methodology}
Our first scenario revealed that the results from a single RAG pipeline can drift over time as new data is added. We now investigate a different dimension of uncertainty: at a single point in time, do different state-of-the-art embedding models produce consistent results for the same query? This scenario quantifies the agreement between models on a downstream retrieval task, measuring how contingent the final search results are on the initial choice of embedding model.

\paragraph{Experimental Protocol}
To isolate the embedding model as the sole variable, we enforced a strictly controlled retrieval environment:
\begin{itemize}
    \item \textbf{Models Tested:} BGE, E5, and Qwen.
    \item \textbf{Index Type:} We used a FAISS \texttt{Flat\_L2} index for all tests, which performs an exact, brute-force search to eliminate any variation from approximate algorithms.
    \item \textbf{Precision:} All models were run using \texttt{FP32} precision.
    \item \textbf{Dataset:} A 5,000-document subset of MSMARCO with 100 queries. Top-50 documents were retrieved.
\end{itemize}
We then performed pairwise comparisons of the top-50 retrieved document lists for each of the three model pairs (BGE vs. E5, BGE vs. Qwen, E5 vs. Qwen) across all 100 queries.

\subsubsection{Results and Discussion}
Our analysis reveals a significant divergence in the retrieval results produced by the different embedding models, underscoring that the choice of model is a dominant source of system-level uncertainty.

\paragraph{Low Agreement Across All Metrics}
As summarized in Table~\ref{tab:cross_model_stability}, the agreement between the models was low. The Overlap Coefficient ranged from just 0.43 to 0.54, meaning that, on average, two state-of-the-art models only agreed on about half of the retrieved documents for the same query. The top-weighted RBO score was similarly moderate, indicating significant disagreement even at the top of the ranked lists.

The most striking result is the very low Kendall's Tau score (0.32–0.38). This indicates that even for the subset of documents that both models did retrieve, their relative ranking was very different. Furthermore, with p-values > 0.05, this weak correlation is not statistically significant.

\begin{table}[h]
  \caption{Mean Retrieval Agreement Between Model Pairs Across 100 Queries.}
  \label{tab:cross_model_stability}
  \begin{tabular}{lccc}
    \toprule
    Model Pair & Overlap Coeff. & RBO Score & Kendall's Tau \\
    \midrule
    \texttt{BGE vs. E5} & 0.540 & 0.570 & 0.384 \\
    \texttt{BGE vs. Qwen} & 0.454 & 0.486 & 0.338 \\
    \texttt{E5 vs. Qwen} & 0.432 & 0.474 & 0.322 \\
    \bottomrule
  \end{tabular}
\end{table}

\begin{figure*}[ht!]
    \centering
    \begin{subfigure}[b]{0.32\textwidth}
        \centering
        \includegraphics[width=\textwidth]{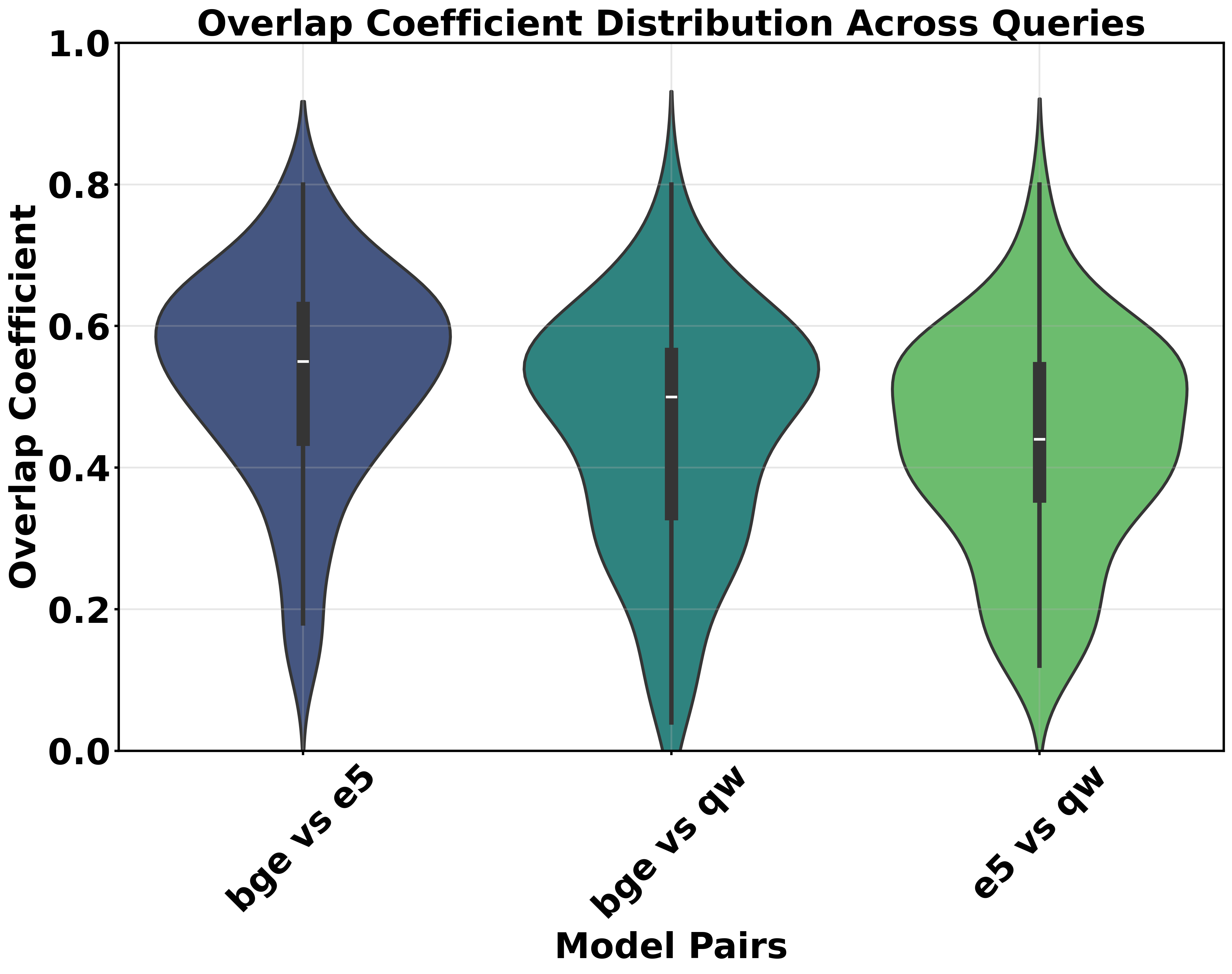}
        \caption{Overlap Coefficient Distribution}
        \label{fig:violin_overlap}
    \end{subfigure}
    \hfill
    \begin{subfigure}[b]{0.32\textwidth}
        \centering
        \includegraphics[width=\textwidth]{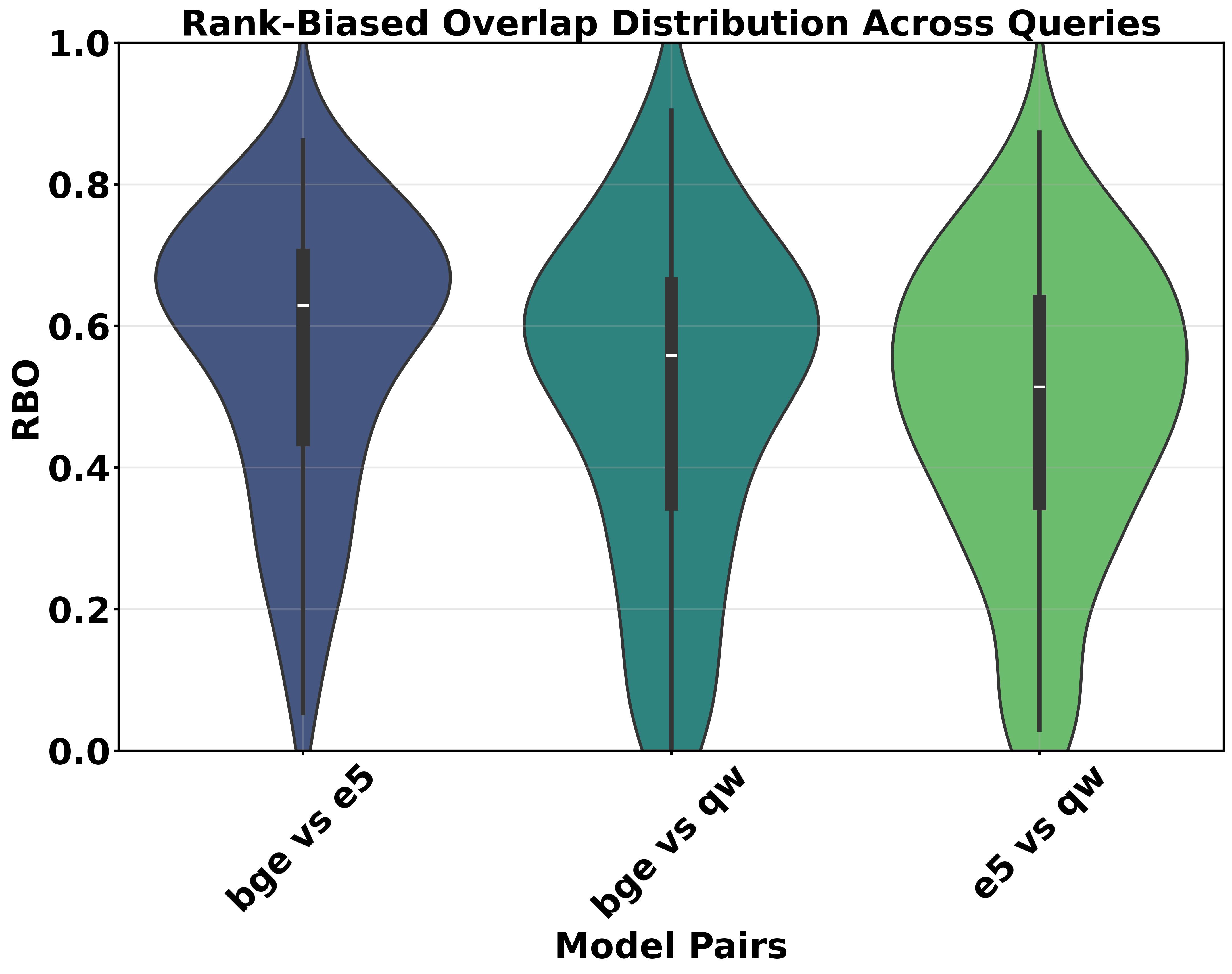}
        \caption{Rank-Biased Overlap (RBO) Distribution}
        \label{fig:violin_rbo}
    \end{subfigure}
    \hfill
    \begin{subfigure}[b]{0.32\textwidth}
        \centering
        \includegraphics[width=\textwidth]{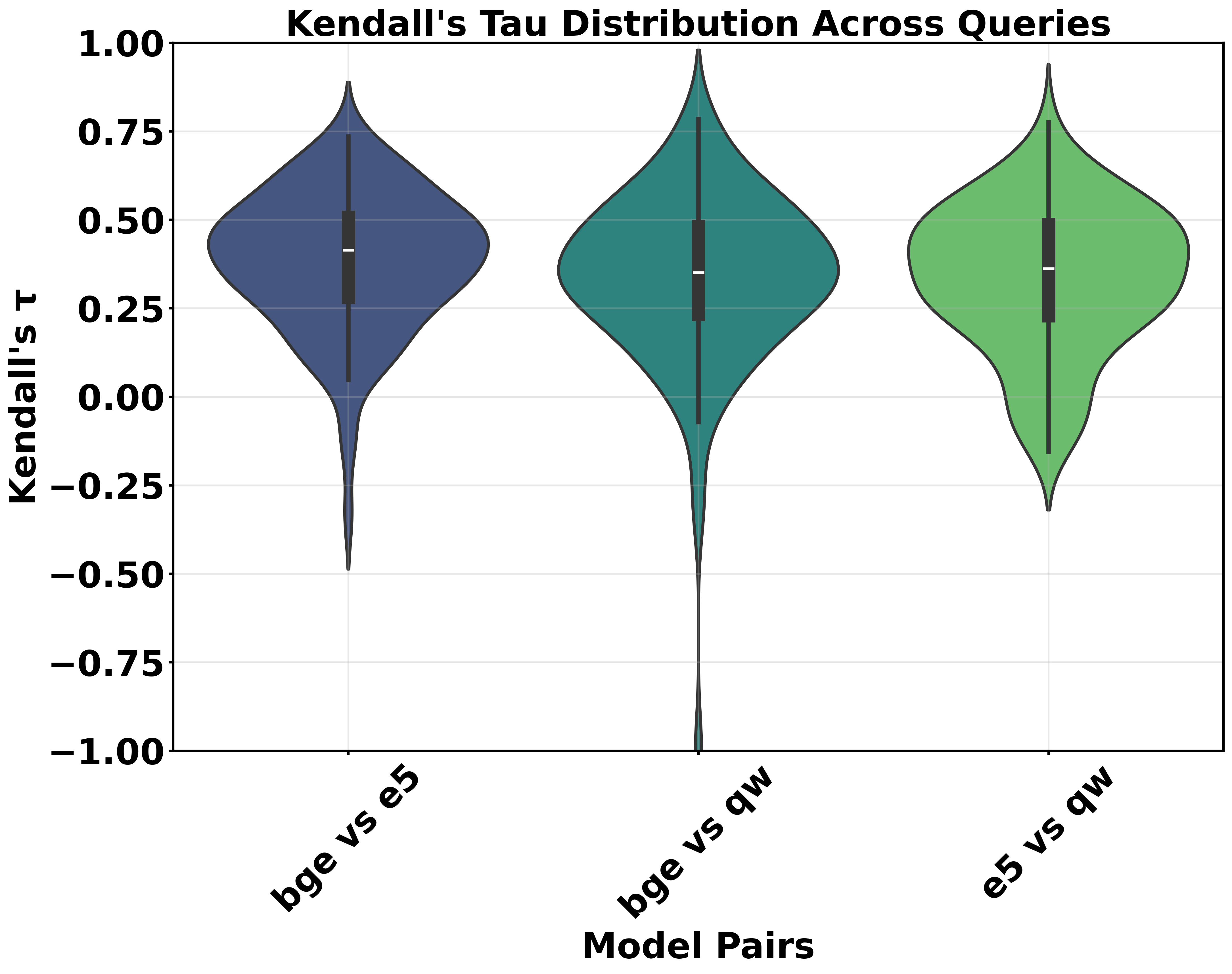}
        \caption{Kendall's Tau Distribution}
        \label{fig:violin_kendall}
    \end{subfigure}
    \caption{Distributions of retrieval agreement metrics for three model pairs across 100 queries. The low median values and wide distributions highlight a significant and variable lack of consensus between the models.}
    \label{fig:cross_model_violin}
\end{figure*}

\paragraph{Visualizing the Distribution of Disagreement}
Figure~\ref{fig:cross_model_violin} visualizes the full distributions of these metrics across the 100 test queries. The violin plots show not only that the median agreement is low but also that there is high variance. For some queries, the models had almost no overlap (RBO scores near 0.2), while for others, they agreed more strongly. This indicates that the disagreement is not uniform but query-dependent.

\paragraph{Discussion}
This scenario provides a critical insight for the "Hierarchy of Uncertainty." \textcolor{blue}{While our other scenarios show that a single RAG pipeline can be made perfectly reproducible, this experiment demonstrates that the results are nonetheless highly contingent on the initial choice of embedding model. This represents a major challenge for the reproducibility of scientific findings that rely on RAG systems. A conclusion drawn using a BGE-based RAG may not be reproducible with an E5-based RAG, as they are fundamentally "seeing" different evidence. This underscores the need for researchers to be explicit about the embedding models used in their work and for the community to develop best practices for evaluating the robustness of findings across different models.}

\subsection{Scenario 3: Precision Uncertainty}
\subsubsection{Goal and Methodology}
We investigate the impact of numerical precision and algorithmic determinism on the stability of the embedding generation process itself, which is a foundational source of uncertainty in any RAG system. The objective is to quantify how variations in floating-point formats and \textit{non-deterministic} (non-det.) settings affect the consistency of vector embeddings before they are used for retrieval.

To achieve this, ReproRAG employs a comprehensive eight-configuration testing framework that examines embedding reproducibility across two key dimensions:
\begin{itemize}
    \item \textbf{Precision Formats:} Four numerical precision types are tested: full-precision \texttt{FP32}, half-precision \texttt{FP16}, and specialized formats \texttt{BF16} and \texttt{TF32} on compatible hardware.
    \item \textbf{Deterministic Modes:} Two algorithmic settings are used: a \textit{deterministic} (det.) mode and a non-det. mode. The \textbf{det. mode} uses fixed random seeds and deterministic cuDNN settings, while the \textbf{non-det. mode} uses time-based seeds and performance-optimized cuDNN settings.
\end{itemize}
This 8-configuration testing matrix (e.g., \texttt{FP32-det.}, \texttt{FP32-non-det.}, etc.) is applied to three widely-used embedding models—BGE, E5, and Qwen—on a diverse set of test texts.

\paragraph{Reproducibility Testing Protocol}
Our methodology consists of two distinct testing protocols:
\begin{enumerate}
    \item \textbf{Same-Configuration Reproducibility:} For each of the eight configurations, we perform five independent embedding generation runs on identical input texts. This protocol tests whether a single, specific setup (e.g., \texttt{FP32-det.}) produces the same embeddings every time it is executed.
    \item \textbf{Cross-Configuration Precision Analysis:} We conduct pairwise comparisons of the generated embeddings \textit{between} different precision formats (e.g., comparing the \texttt{FP32-det.} output to the \texttt{FP16-det.} output). This quantifies the vector drift caused purely by changes in numerical precision.
\end{enumerate}

\paragraph{Technical Implementation}
To ensure the integrity of the experiment, ReproRAG enforces strict environment controls:
\begin{itemize}
    \item In \textbf{det. mode}, we fix random seeds across all relevant libraries (\texttt{torch.manual\_seed(42)}, \texttt{np.random.seed(42)}) and enable deterministic cuDNN operations (\texttt{cudnn.\\deterministic=True}, \texttt{cudnn.benchmark=False}).
    \item In \textbf{non-det. mode}, we use time-based seeds and optimized, non-deterministic cuDNN settings (\texttt{cudnn.benchmark=True}) to reflect a production performance-oriented environment.
    \item To prevent state leakage, a fresh model instance is loaded for each experimental run.
\end{itemize}

\subsubsection{Results and Discussion}
Our evaluation of embedding uncertainty was conducted on three widely-used transformer-based models: BGE, E5, and Qwen. The analysis yielded two clear and impactful findings. First, and most surprisingly, all three models were perfectly reproducible across all tested configurations, and the software-level determinism flags had no observable effect on their output. Second, the choice of numerical precision, while a definite source of variation, results in a quantitatively small ``embedding drift''.

\paragraph{The Ineffectiveness of Determinism Flags}
Our primary goal was to measure the impact of common determinism controls.We tested eight configurations for each model—four precision formats, each in both a \textit{deterministic} (det.) and a \textit{non-deterministic} (non-det.) mode. The results were uniform.

While all three models produced identical outcomes, we present the detailed results for the BGE model as a representative example.
\begin{enumerate}
    \item \textbf{Perfect Same-Configuration Reproducibility:} As shown in Table~\ref{tab:embedding_intra_repro}, Every single one of the eight configurations was perfectly reproducible over five independent runs, achieving a mean pairwise L2 distance of 0.0 and cosine similarity of 1.0.
    \item \textbf{Identical det. vs. non-det. Outputs:} As shown in Table~\ref{tab:det_vs_nondet}, for every precision format, the embeddings generated in det. mode were numerically identical to those generated in non-det. mode.
\end{enumerate}

 As summarized in Table~\ref{tab:embedding_intra_repro_summary}, every single configuration for all three models was perfectly reproducible over five independent runs. This leads to a critical insight: \textcolor{blue}{The deterministic behavior observed is not specific to one model but may be a general property of modern, optimized transformer embedding models on a standard HPC software stack. The \texttt{cudnn.deterministic} and \texttt{cudnn.benchmark} flags did not influence the final embedding vectors.} This result demonstrates that practitioners cannot assume that these flags will induce or control randomness; reproducibility must be empirically verified.

\begin{table}[h]
  \caption{Same-Configuration Embedding Reproducibility over 5 runs. An L2 distance of 0 indicates numerically identical vectors, and a cosine similarity of 1 indicates perfect angular agreement. All eight configurations achieved perfect reproducibility.}
  \label{tab:embedding_intra_repro}
  \begin{tabular}{lccc}
    \toprule
    Configuration  & L2 Distance &Cosine Similarity & Reproducible \\
    \midrule
    \texttt{FP32-det.}  & \texttt{0.00} &\texttt{1.00} &Yes \\
    \texttt{FP32-non-det.}  & \texttt{0.00} &\texttt{1.00} & Yes \\
    \texttt{FP16-det.}  & \texttt{0.00} &\texttt{1.00} & Yes \\
    \texttt{FP16-non-det.} & \texttt{0.00} &\texttt{1.00} & Yes \\
    \texttt{BF16-det.}  & \texttt{0.00} &\texttt{1.00} & Yes \\
    \texttt{BF15-non-det.}  & \texttt{0.00} &\texttt{1.00} & Yes \\
    \texttt{TF32-det.}  & \texttt{0.00} &\texttt{1.00} & Yes \\
    \texttt{TF32-non-det.}  & \texttt{0.00} &\texttt{1.00} & Yes \\
    \bottomrule
  \end{tabular}
\end{table}

\begin{table}[h]
  \caption{Summary of Intra-Configuration Reproducibility Across Three Models. Each model was tested over 8 configurations (4 precisions x 2 determinism modes), with 5 runs per configuration.}
  \label{tab:embedding_intra_repro_summary}
  \begin{tabular}{lcc}
    \toprule
    Embedding Model & Reproducible Configurations  \\
    \midrule
    \texttt{BGE} & 8 out of 8  \\
    \texttt{E5} & 8 out of 8  \\
    \texttt{Qwen} & 8 out of 8 \\
    \bottomrule
  \end{tabular}
\end{table}

\begin{figure}[h]
  \centering
  \includegraphics[width=0.9\columnwidth]{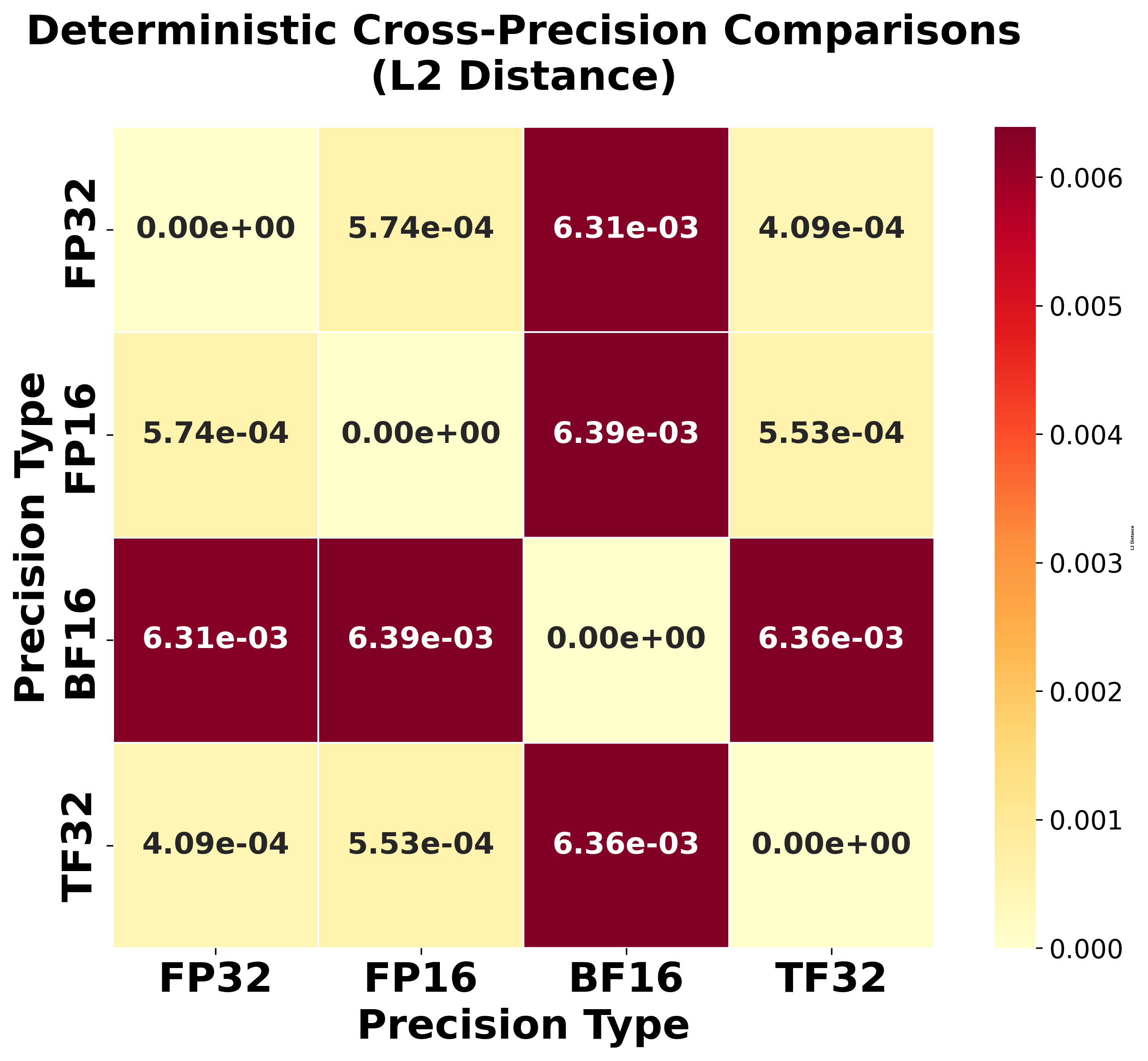}
  \caption{Heatmap of pairwise L2 distances between precision formats in Deterministic mode. Darker colors indicate greater dissimilarity.}
  \label{fig:heatmap_det}
\end{figure}

\paragraph{Quantifying Precision-Induced Embedding Drift}
While software flags had no effect, the choice of numerical precision itself was a guaranteed source of variation. Our cross-configuration analysis confirmed that using different floating-point formats fundamentally alters the resulting embeddings. Figure~\ref{fig:heatmap_det} visualizes the pairwise L2 distances between the four precision formats.

The heatmap clearly shows that all off-diagonal comparisons result in non-zero distances. However, the magnitude of this drift is very small. For instance, the distance between the \texttt{FP32} baseline and \texttt{FP16} is only \texttt{5.74e-04}. This analysis allows us to rank the precision formats by their similarity. \texttt{TF32} is most similar to the \texttt{FP32} baseline (L2 distance of \texttt{4.09e-04}), making it a strong candidate for optimization. Conversely, \texttt{BF16} is the most distinct, exhibiting the largest drift from all other formats (e.g., an L2 distance of \texttt{6.31e-03} from \texttt{FP32}).

This analysis provides two key takeaways. \textcolor{blue}{First, changing precision will cause embedding drift. Second, this drift, while real, is quantitatively small for most pairs.} This allows practitioners to make informed trade-offs between computational efficiency and numerical fidelity, armed with the knowledge of how much the numerical embedding vector space will be perturbed.

\begin{figure*}[ht!]
    \centering
    \begin{subfigure}[b]{0.48\textwidth}
        \centering
        \includegraphics[width=\textwidth]{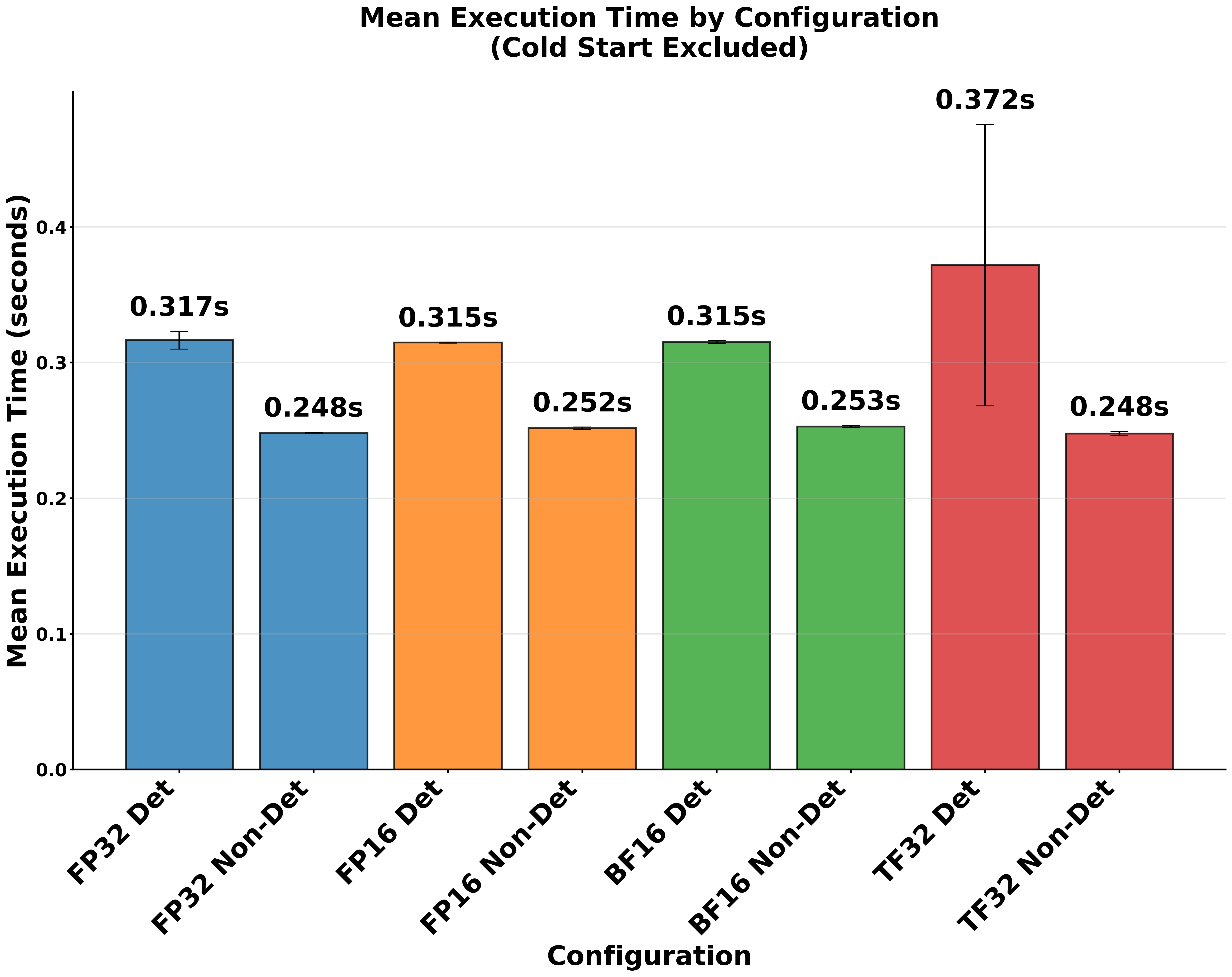}
        \caption{BGE Model Latency}
        \label{fig:bge_latency}
    \end{subfigure}
    \hfill 
    \begin{subfigure}[b]{0.48\textwidth}
        \centering
        \includegraphics[width=\textwidth]{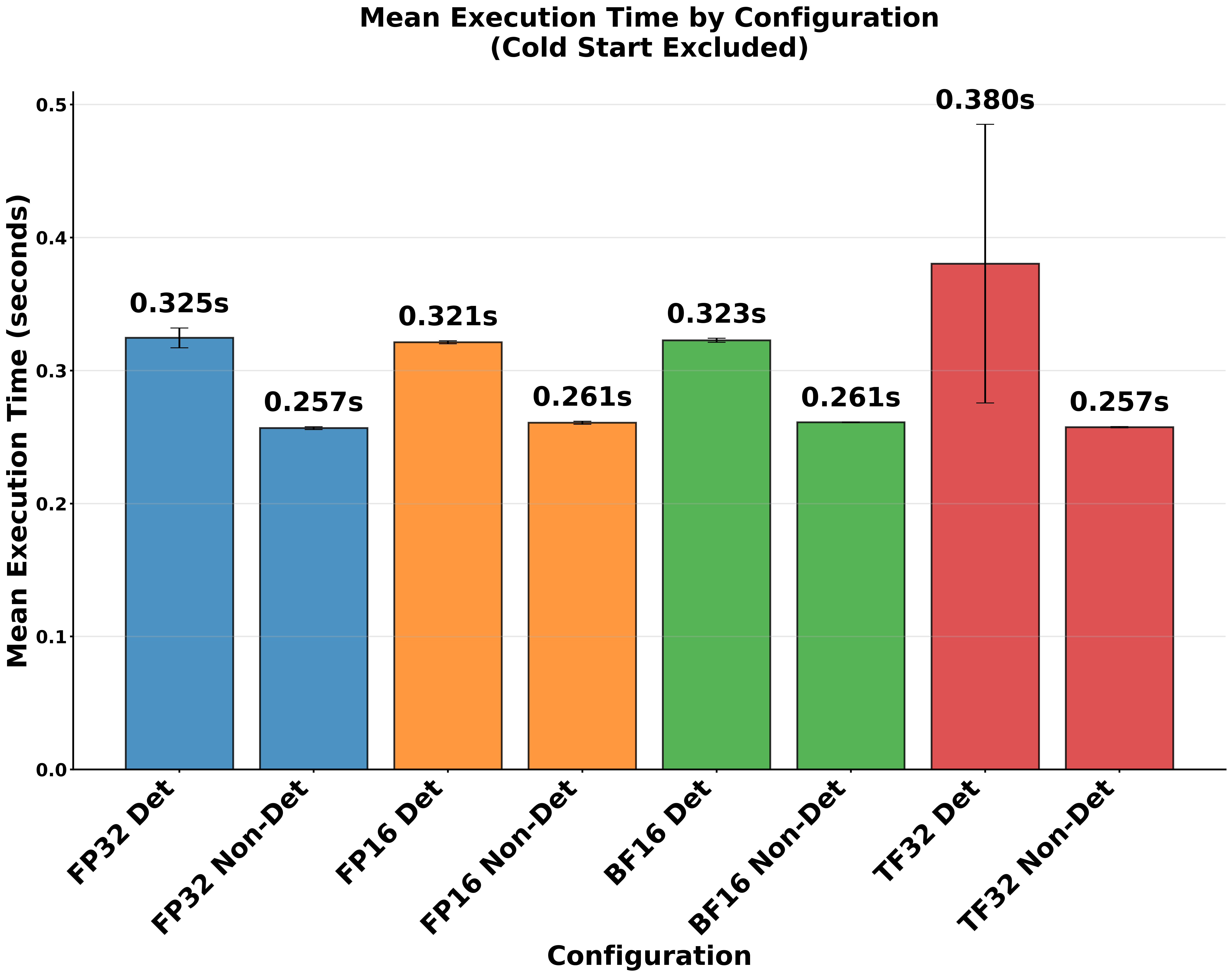}
        \caption{E5 Model Latency}
        \label{fig:e5_latency}
    \end{subfigure}
    \hfill 
    \begin{subfigure}[b]{0.48\textwidth}
        \centering
        \includegraphics[width=\textwidth]{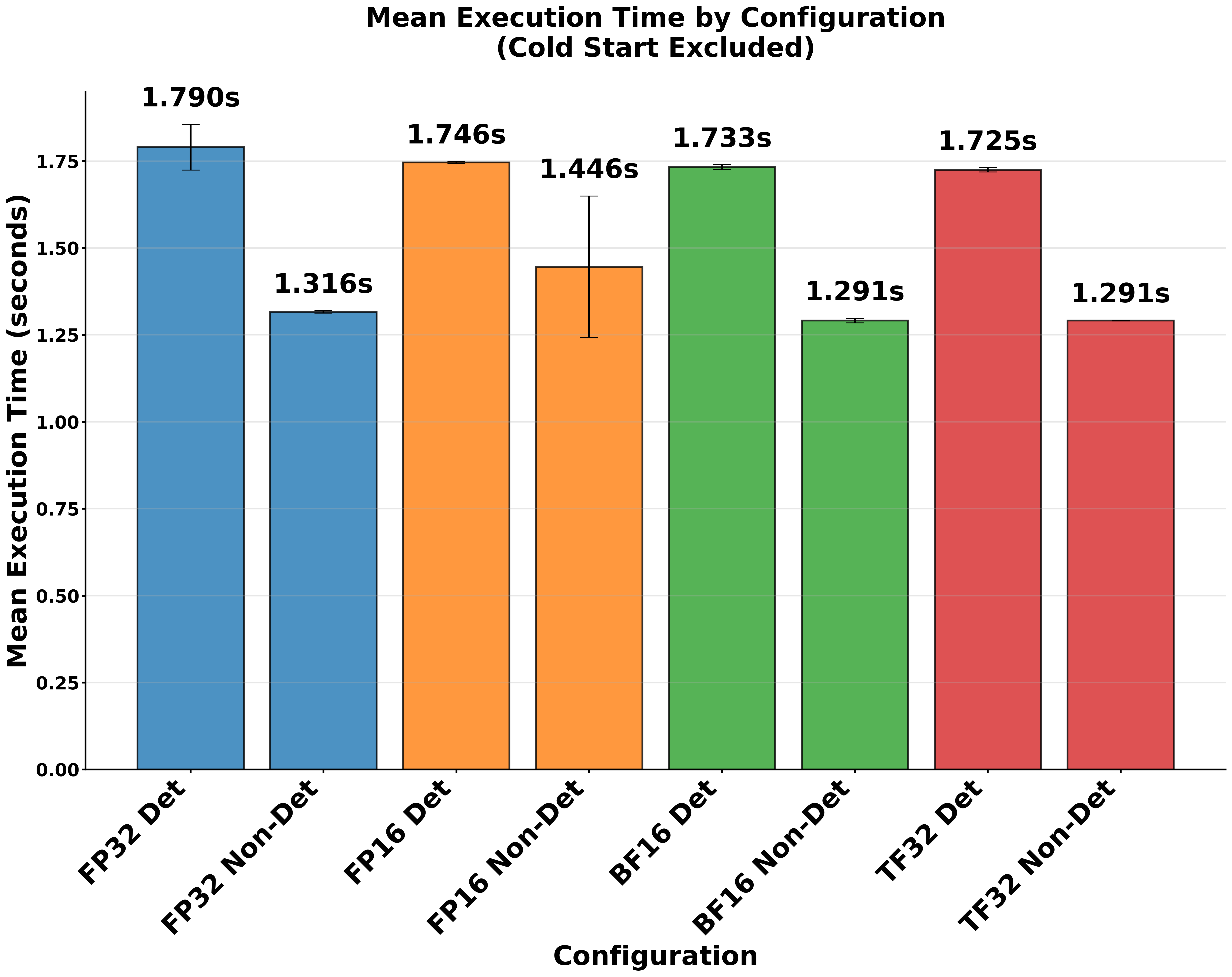}
        \caption{Qwen Model Latency}
        \label{fig:qwen_latency}
    \end{subfigure}

    \caption{Embedding generation latency for each model across all 8 configurations, spanning the full page width for clarity. Lower is better. The non-deterministic mode is consistently faster across all models and precisions.}
    \label{fig:all_latency}
\end{figure*}

\paragraph{Performance Analysis}
To complete our analysis of the embedding stage, we measured the execution time for all 24 scenarios (3 models x 8 configurations). Figure~\ref{fig:all_latency} shows the embedding generation latency for each model. The results reveal a clear and actionable insight regarding the determinism settings.

Across all three models and all precision formats, we observed a consistent performance trend: the \textit{non-deterministic} (non-det.) mode was measurably faster than the \textit{deterministic} (det.) mode, typically by a margin of \texttt{10-15\%}.

This creates a powerful takeaway when combined with our earlier findings. Our reproducibility analysis showed that for these models, the det. and non-det. modes produced numerically identical embeddings. \textcolor{blue}{This performance analysis now demonstrates that the non-det. mode is not only harmless to reproducibility but is also faster. This means practitioners can enable the performance-oriented cuDNN settings to gain a ``free'' and consistent speedup without sacrificing any output reproducibility for this class of models.}

\begin{table}[h]
  \caption{Comparison of Deterministic between Non-Deterministic Modes Within Each Precision Format.}
  \label{tab:det_vs_nondet}
  \begin{tabular}{lcc}
    \toprule
    Precision Type & L2 Distance  & Result \\
    \midrule
    \texttt{FP32} & \texttt{0.00} & Identical \\
    \texttt{FP16} & \texttt{0.00} & Identical \\
    \texttt{BF16} & \texttt{0.00} & Identical \\
    \texttt{TF32} & \texttt{0.00} & Identical \\
    \bottomrule
  \end{tabular}
\end{table}

\subsection{Scenario 4: Stability of the Indexing and Search Stage}
\subsubsection{Goal and Methodology}
The index uncertainty experiment aims to evaluate the reproducibility of vector indexing and search operations, isolating this critical stage from the rest of the RAG pipeline. This investigation addresses whether identical document embeddings consistently produce the same retrieval behaviors across multiple runs. The primary goals are to (1) Quantify the deterministic behavior over different FAISS indexing techniques, (2) Assess the impact of GPU-accelerated \textit{search} on retrieval consistency, and (3) Measure how indexing variations propagate to final retrieval results.

To achieve this, ReproRAG employs a multi-layered testing approach using a controlled environment.
\paragraph{Controlled Environment and Configurations}

Each test run is initialized with identical random seeds (42) when \textit{deterministic} (det.) mode is enabled. The framework also supports a \textit{non-deterministic} (non-det.) mode using time-based seeds for sensitivity analysis. The evaluation covers multiple FAISS indexing techniques, including \texttt{IndexFlatL2} (exact), \texttt{IndexIVFFlat} (inverted file), and \texttt{IndexHNSWFlat} (hierarchical navigable small world). 
For GPU testing, a CPU-built index is transferred to a GPU resource for accelerated search operations. Note that, to isolate the retrieval stage, all test runs for this scenario utilize identical, pre-computed, and cached document embeddings.

\paragraph{Reproducibility Measurement Protocol}
For each index configuration, the system performs five independent runs. Each run consists of building a FAISS index from the cached embeddings and then executing a standard set of queries against it. For GPU-specific tests, this involves transferring the index to the GPU before querying. We measure the end-to-end retrieval consistency using our core metrics (EMR, Jaccard Similarity, Kendall's Tau). In addition, for clustering-based indices like IVF, we perform a \textbf{centroid stability analysis} by measuring the L2 distance between the cluster centroids generated in different runs.

\subsubsection{Results and Discussion}
The results of our retrieval uncertainty evaluation were definitive and surprising: we observed perfect reproducibility across every tested configuration.

\paragraph{Uniform Stability Across All Index Types}
As shown in Table~\ref{tab:retrieval_repro}, all tested index types—from brute-force \texttt{Flat\_L2} to complex approximate methods like \texttt{HNSW}—achieved a perfect 1.000 score on all reproducibility metrics when run on the CPU. The Exact Match Rate, Jaccard Similarity, and Kendall's Tau were all perfect, indicating that identical document rankings were produced in every run. This finding demonstrates that with modern libraries like FAISS and under a strictly controlled environment (i.e., with fixed random seeds), the indexing and CPU search algorithms themselves behave in a perfectly reproducible manner.

\begin{table}[h]
  \caption{Retrieval Reproducibility Across Various FAISS Index Types on CPU. All configurations demonstrated perfect stability.}
  \label{tab:retrieval_repro}
  \begin{tabular}{lccc}
    \toprule
    Index Type & Exact Match & Jaccard & Kendall Tau \\
    \midrule
    \texttt{Flat\_L2} & 1.00 & 1.00 & 1.00 \\
    \texttt{Flat\_IP} & 1.00 & 1.00 & 1.00 \\
    \texttt{IVF} & 1.00 & 1.00 & 1.00 \\
    \texttt{HNSW\_accurate} & 1.00 & 1.00 & 1.00 \\
    \texttt{HNSW\_fast} & 1.00 & 1.00 & 1.00 \\
    \texttt{LSH} & 1.00 & 1.00 & 1.00 \\
    \bottomrule
  \end{tabular}
\end{table}

\paragraph{GPU Search Operations Had No Impact on Reproducibility}
Furthermore, our targeted tests of GPU-accelerated search yielded the same result. \textcolor{blue}{After transferring a deterministically built index to the GPU, the search results remained perfectly reproducible. Toggling determinism controls for GPU operations during the search phase had no impact on the final retrieved document list; all runs remained identical.} 

\paragraph{Discussion}
\textcolor{blue}{The key insight from this scenario is that the FAISS retrieval stage—both indexing and search on CPU or GPU—when isolated and properly controlled, is not a significant source of non-determinism.} This is a crucial finding for practitioners. It implies that when debugging reproducibility issues in a RAG system, the core FAISS operations are unlikely to be the culprit, provided that environmental factors like random seeds are managed. This shifts the focus of concern towards other parts of the pipeline, such as the embedding drift we quantified in the previous section, data versioning, or inconsistencies in the broader software environment. Our framework, by allowing us to isolate this component, has successfully falsified a common hypothesis about a primary source of system instability.

\subsection{Scenario 5: Distributed Uncertainty}

\subsubsection{Goal and Methodology}
This scenario evaluates reproducibility in a multi-node HPC environment, the setting where non-determinism is often most pronounced due to network timing, race conditions, and distributed coordination challenges. The goal is to measure the end-to-end consistency of our distributed retrieval protocol.

\paragraph{System Architecture and Protocol}
Our distributed system implements a shared-nothing, scatter-gather architecture using MPI for inter-node communication. The key phases of the protocol are:
\begin{enumerate}
    \item \textbf{Document Sharding:} The document collection is partitioned across all available nodes. We test three distinct distribution strategies: \textbf{hash-based} (using \texttt{hash(doc\_id) \% num\_nodes} for deterministic balance), \textbf{range-based} (sequential blocks per node), and \textbf{random} (with a fixed seed for run-to-run consistency).
    \item \textbf{Parallel Indexing:} Each node builds its local FAISS index independently using only its assigned document shard. This phase is ``embarrassingly parallel'' and requires no inter-node communication. An MPI barrier ensures all nodes complete indexing before proceeding.
    \item \textbf{Distributed Search and Aggregation:} A query is broadcast to all nodes. Each node searches its local index in parallel. The top-k candidate results from all nodes are gathered to a root process (rank 0).
    \item \textbf{Global Aggregation:} The root process performs a global merge of all candidate documents, re-ranks them based on their true similarity scores (e.g., L2 distance), and selects the final global top-k results.
\end{enumerate}

\paragraph{Experimental Configurations}
We conducted these tests on a 4-node cluster within TAMU ACES. We evaluated the reproducibility of the distributed protocol across all the FAISS index types previously tested in the single-node scenario.

\subsubsection{Results and Discussion}
Across all tested configurations, the distributed retrieval protocol demonstrated perfect, 100\% reproducibility.

\paragraph{Perfect Consistency Across All Factors}
As summarized in Table~\ref{tab:distributed_repro}, every combination of index type and sharding strategy yielded perfect scores on all our reproducibility metrics. The Exact Match Rate, Jaccard Similarity, and Kendall's Tau were all 1.000, indicating that identical, identically-ranked results were returned in every run. This result held true for both simple indexes like \texttt{Flat\_L2} and complex approximate indexes like \texttt{HNSW}.

\begin{table}[h]
  \caption{Distributed Retrieval Reproducibility (4 Nodes). For all metrics, a score of 1.0 indicates perfect reproducibility (the best possible score), while 0.0 would indicate no overlap or correlation (the worst possible score).}
  \label{tab:distributed_repro}
  \begin{tabular}{lcccc}
    \toprule
    Index Type &Sharding Type &Exact Match & Jaccard & Kendall Tau \\
    \midrule
    \texttt{Flat\_L2} & Hash & 1.00 & 1.00 & 1.00 \\
    \texttt{IVF} & Hash & 1.00 & 1.00 & 1.00 \\
    \texttt{HNSW} & Hash & 1.00 & 1.00 & 1.00 \\
    \texttt{LSH} & Hash & 1.00 & 1.00 & 1.00 \\
    \texttt{Flat\_L2} & Range & 1.00 & 1.00 & 1.00 \\
    \texttt{IVF} & Range & 1.00 & 1.00 & 1.00 \\
    \texttt{HNSW} & Range & 1.00 & 1.00 & 1.00 \\
    \texttt{LSH} & Range & 1.00 & 1.00 & 1.00 \\
        \texttt{Flat\_L2} &Random & 1.00 & 1.00 & 1.00 \\
    \texttt{IVF} & Random & 1.00 & 1.00 & 1.00 \\
    \texttt{HNSW} & Random & 1.00 & 1.00 & 1.00 \\
    \texttt{LSH} & Random & 1.00 & 1.00 & 1.00 \\
    \bottomrule
  \end{tabular}
\end{table}

\paragraph{Reproducibility by Design}
This perfect stability is not accidental but a direct result of the protocol's design. The key factors ensuring reproducibility are:
\begin{itemize}
    \item \textbf{Deterministic Sharding:} Using hash-based or seeded-random sharding ensures that every node receives the exact same subset of documents on every run.
    \item \textbf{MPI Synchronization:} The use of MPI barriers guarantees that all nodes complete critical phases (like indexing) before any node can advance, eliminating temporal race conditions.
    \item \textbf{Centralized Aggregation:} The gather-and-aggregate step on a single root node ensures that the final global ranking logic is performed in a serial, deterministic manner.
\end{itemize}

\paragraph{Discussion}
\textcolor{blue}{The key insight from this scenario is that the complexity of distributed operations does not inherently lead to non-determinism, provided the communication and data partitioning protocol is carefully designed for reproducibility.} While the system has typical distributed computing trade-offs—such as network overhead during the gather phase and a temporary memory concentration on the root node—our results prove that it does not sacrifice consistency. This validation is critical for deploying RAG systems in production scientific environments where bit-for-bit identical results are often required. Our framework has successfully demonstrated that this distributed architecture is suitable for such high-stakes applications.

\section{Conclusion}

This paper introduced ReproRAG, a comprehensive framework for benchmarking the reproducibility of RAG retrieval systems. Our empirical evaluation, conducted across a series of controlled scenarios, not only quantifies sources of non-determinism but also challenges several common assumptions held by the community.

Our findings reveal a clear hierarchy of reproducibility challenges. While data insertion causes predictable result displacement (Scenario 1), and precision choice causes measurable drift (Scenario 2), we found that the core indexing/search algorithms (Scenario 3) and distributed protocols (Scenario 4) are, under proper control, perfectly stable on a run-to-run basis. This hierarchy suggests that efforts to improve RAG reproducibility should focus less on the algorithms themselves and more on data versioning, precision management, and strategies for handling dynamic data.

A key contribution of this work is the empirical falsification of the common hypothesis that approximate nearest neighbor algorithms are a primary source of run-to-run randomness. Similarly, we demonstrated that common software flags (\texttt{cudnn.deterministic}) may have no observable effect on the output of modern embedding models, underscoring the necessity of empirical validation over reliance on framework settings.

In conclusion, this work provides both the tools and the empirical evidence to move the discussion of RAG reproducibility from anecdote to quantitative measurement. The open-source ReproRAG framework empowers users to validate their own systems and make informed architectural decisions. Future work will involve extending the framework to other popular vector databases and integrating it into CI/CD pipelines for the continuous validation of trustworthy AI in science.

\begin{acks}
This work is based upon support of the Summer of Reproducibility, a program funded by the US National Science Foundation under Grant No. 2226407.
This work used TAMU ACES at Texas A\&M University through allocation CHE240191 from the Advanced Cyberinfrastructure Coordination Ecosystem: Services \& Support (ACCESS) program, which is supported by National Science Foundation grants \# 2138259, \# 2138286, \# 2138307, \# 2137603 and \# 2138296. 
This research is supported in part by the U.S.\@ Department of Energy (DOE) through
  the Office of Advanced Scientific Computing Research's ``Orchestration for Distributed \& Data-Intensive Scientific Exploration'' and
  the ``Cloud, HPC, and Edge for Science and Security'' LDRD at Pacific Northwest National Laboratory.
PNNL 
is operated by Battelle for the DOE under Contract DE-AC05-76RL01830.
\end{acks}

\bibliographystyle{ACM-Reference-Format}
\bibliography{references}

\end{document}